# The design of China Reconfigurable Analog-digital backEnd for FAST

Xin-xin Zhang[1], Ran Duan[1], Xin-ying Yu[2], Di Li[1], Ning-yu Tang[1] and Dao-chong Qing[1]

[1] National Astronomical Observatories, Chinese Academy of Sciences, Beijing 100012, China; zhangxinxin@nao.cas.cn

[2] Guizhou University, Guiyang 550025, China;

**Abstract** The Five-hundred-meter Aperture Spherical radio Telescope(FAST) was launched on September 25,2016.From early 2017,we began to use the FAST wideband receiver,which was designed,constructed and installed on the FAST in Guizhou,China.The front end of the receiver is composed an uncooled Quad Ridge Flared Horn feed(QRFH) with the frequency range of 270 to 1620 MHz,and a cryostat operating at 10 K.Stephen et al. 2016We have coop-erated with the Institute of Automation of the Chinese Academy of Sciences to developed the China Reconfigurable ANalog-digital backEnd.The system covers the 3 GHz operating band of FAST.The hardware part of the backend includes an Analog Front-end Board,a wideband high precision Analog Digital Converter,and a FAST Digital Back-end.Analog circuit boards, field programmable gate arrays, and control computers form a set of hardware, software, and firmware platforms to achieve flexible bandwidth requirements through parameter changes. It is also suitable for the versatility of different astronomical observations, and can meet specific requirements. This paper briefly introduces the hardware and software of CRANE, as well as some observations of the system.

## 1 INTRODUCTION

China Reconfigurable Analog-digital backEnd Xia et al. (2016) consists of an Analog Front-end Board,a wideband high precision Analog Digital Converter,and a FAST Digital Back-end.Since the feed cabin is too far away from the control room,the electrical signal from the feed cabin is first converted into an optical signal through the electro-optic converter,and then transmitted to the control room via the opti-cal fiber.Afterwards,it is converted into an electrical signal by the photoelectric converter in the control room.After amplification on the AFB board,it is adjusted to the appropriate power range of the wideband high precision Analog Digital Converter.Finally,the signal is converted into spectrum by FDB.Although the signal is transmitted to the control room via fiber optics,it is still weak.The operating power range of the wideband high precision Analog Digital Converter input signal is between - 2 dBm and 2 dBm.Therefore,the gain of each stage amplifier should not exceed the P1dB output power of the next stage amplifier,otherwise it can not work properly.Also the final SNR should be as large as possible,to guarantee the input power of ADC.It is necessary to consider the selection and ingenious combination of the amplifiers and variable digital attenuators in this AFB system.Two pieces of 3 GHz,12 bit high performance ADC are interleaved to cover the 3 GHz bandwidth.If the Fast Fourier Transform is directly performed on the sampled signal,the problems of spectrum leakage and scalloping loss will occur.The spectrum leakage is related to the sampling frequency and the number of FFTs.The scalloping loss is mainly caused by the characteristics of the single frequency response.In the proposed FDB system, the spectrum leakage of the FFT is mainly suppressed by the polyphase filter bank before its transformation. In the meanwhile,the cascaded FFT algorithm is used in the FDB system, and the 8 M channel FFT is achieved. Through the effective cooperation of the three parts,flexible astronomical observations within 3 GHz bandwidth can be achieved.

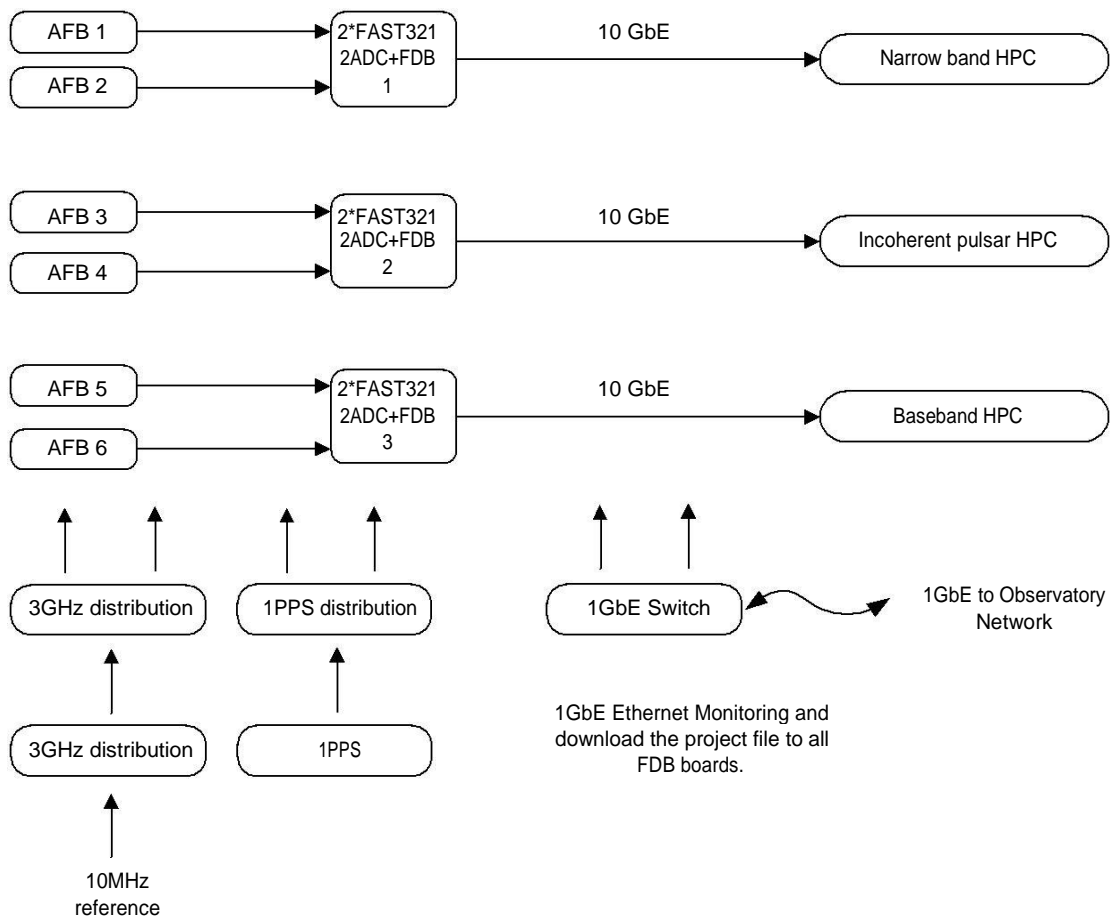

Fig. 1: The system diagram of CRANE.

## 2 ANALOG FRONT-END BOARD

**2.1 Function and performance**

The Analog Front-end Board aims to provide a stable external clock for Analog Digital Converter,and to adjust the Radio Frequency signal power of the receiver by combining amplifiers and digital variable attenuators.The AFB hardware includes the RF amplifiers,Voltage Controlled Oscillators,mixer,filters and the corresponding matching capacitors and resistors.The AFB system is composed of three paths:The first path consists of low noise amplifiers,fixed attenuators,digital electronic resistance attenuators,splitters,and corresponding matching resistance capacitance circuit.It preliminarily processes the RF signal.Because of its sensitivity to the ultra-wideband feed,the output signal power within the working range of Analog Digital Converter.In the second path,mixing and filtering are added.The Voltage Controlled Oscillator is utilized to provide a local Oscillator for the Mixer,and the corresponding baseband output is selected by the digital switch.In the third path,the parameters of the Voltage Controlled Oscillator are modified to provide a reliable and adjustable clock.

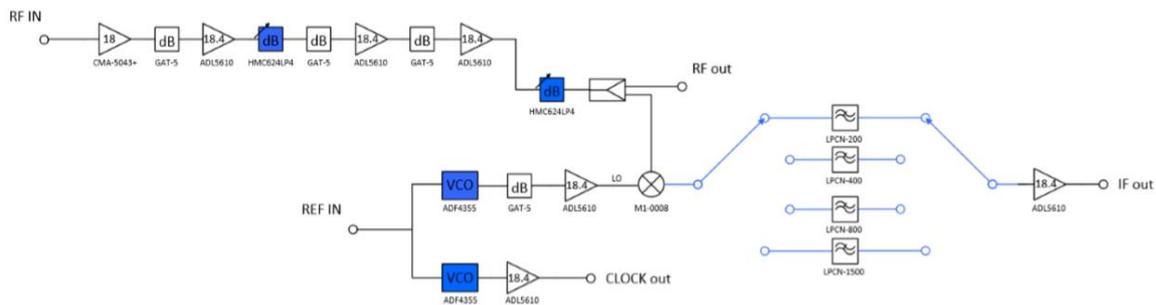

Fig. 2: The system diagram of Analog Front-end Board.

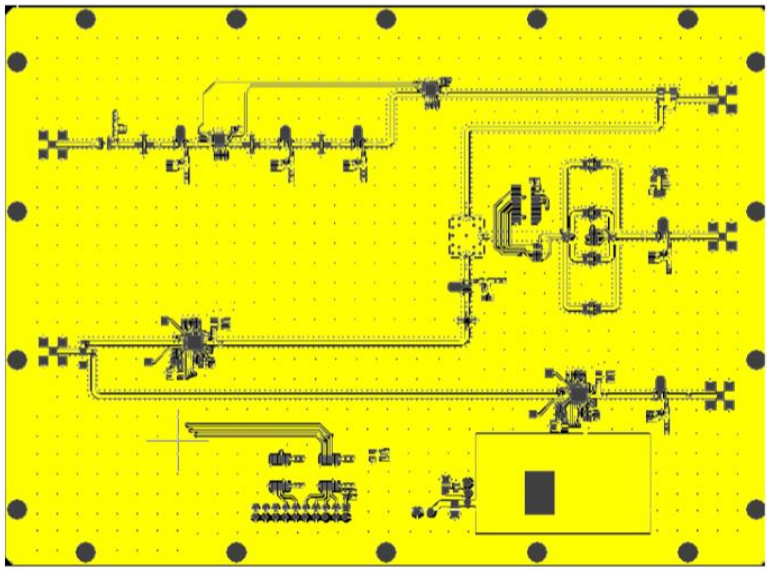

Fig. 3: The Gerber figure of Analog Front-end Board.

To calculate the performance of the AFB system,the input power range of the astronomical signal is -100 dBm - -70 dBm.An Excel calculation table is used to calculate the Signal-to-Noise Ratio of the whole system according to the performance of each RF devices working in this frequency band.The reduction is 0.44 dB and 0.78 dB in 3 GHz,both of which are in the 2 dB range. Many integrated circuit(IC) components are used in the AFB.The intermediate frequency(IF) board can integrate all electronic components between the feed cabin and the FDB.Ran et al. (2015) We designed and fabricated the AFB in 2016.The schematic circuit diagram of the generation of the Gerber file is shown in Fig.3.Fig.4 shows that the fabrication con-tains the following parts:

- Three amplifiers.To select the amplifier with low noise,it is necessary to considerate gain,Noise Figure(NF),third-order intercept point(IP3),P1dB output power,etc.Comprehensively considering all as-pects of device performance,the wideband amplifier CMA-5043+ manufactured by Mini-circuits is se-lected.Since the primary amplifier of the circuit board controls the total noise temperature of the entire IF receiving system,a low-noise,high-gain amplifier is selected,the gain range of which is 10.2 dB - 25.2 dB within 3 GHz bandwidth.The NF range is from 0.73 dB to 1.1 dB,which is about 1 dB smaller than other devices in the same band.The higher Output IP3 has a range 31 dBm to 33.6 dBm and the P1dB range 18.9 dBm to 21.2 dBm.The broadband RF amplifier ADL5610 is selected for three-stage amplification of the signal.

- Three fixed attenuators.The three fixed attenuators are used to suppress the standing wave between the upper stage amplifier and the current stage amplifier,and select the working frequency band from 0.03 GHz to 6 GHz.

- Two Voltage Controlled Oscillators(VCO).One oscillator sets the frequency of Local Oscillator(LO),and the other sets the clock output.The frequency range is set as 137.5 MHz - 4400 MHz.

- Two digital variable attenuators,each of which has the attenuation from 0 to 31.5 dB in 0.5 steps. All functions of the AFB system can be digitally controlled or programmed by a computer.An Arduino board is connected to the AFB via a 20-pin ribbon cable.A python script is developed on the control computer to select the various functions via the visual interface.

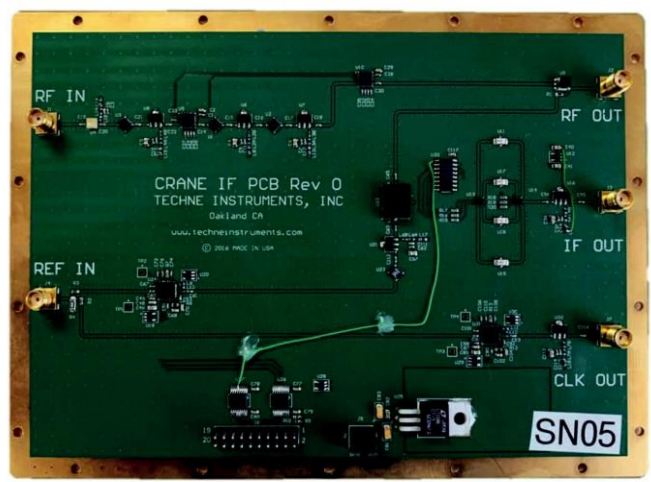

Fig. 4: Picture of real product.

**2.2 Tests**

Firstly,the variation of gain is tested with the 70 MHz-3 GHz passband,and the RF-in port of the AFB is used to input the signals.The input signal power is about -34 dBm.The output signal power at the RF out port is tested.A total of 19 frequency points are tested,the interval of which is about 200 MHz,and the curve is indicated by the blue line.The red line shows the path gain curve after attenuation 10 dB using a digital attenuator.When the ultra-wideband receiver is used for observations,the working bandwidth of ultra-wideband receiver is 270 MHz-1620 MHz.The output gain of signal power is stable at about 45 dB.The multi-beam receiver operates from 1.05 GHz to 1.45 GHz,Stephanie et al. (2017)and the overall gain of the entire band is also maintained at around 45 dB.The receiver system engineers will install the s-band receiver in the future,and the s-band feed operates from 2 GHz to 3 GHz.The band with a gain of 40 dB is tested.The accurate calculations of the previous Excel table ensure that the gain of each amplifier is proper.In this way,its output power is greater than the P1dB output power of the next amplifier,making it saturated and unable to work properly.The AFB ensures that the signal transmitted by the board work will operate normally in the operating band without saturation.

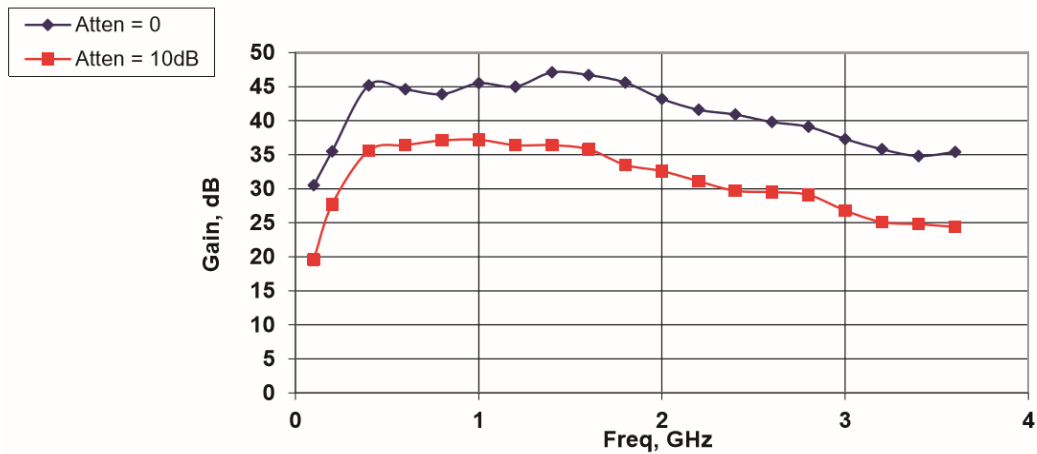

Fig. 5: Frequency response diagram of gain.

The variation of the mixing path gain with different LO frequencies is tested.The initial LO frequency is set to 0.5 GHz,and the input signal of RF-in port of the AFB is 50 higher than the LO setting frequency of 50 MHz.In other words,the IF-out port output signal is stable at 50 MHz.The LO frequency increases gradually.The input signal frequency of the RF-in port is always higher than the LO setting frequency of 50 MHz.A total of 13 frequency points are tested,covering the whole 3 GHz bandwidth.The reason for selecting the IF-out port to output 50 MHz frequency signal is that the attenuation is fixed at 5 dB by querying the data table of four filters corresponding to each path and the attenuation does not fluctuate.From the following figure, it can be found that the overall gain of the AFB changes after the frequency band is down-converted to baseband. Therefore,it is necessary to test the fluctuation state of the gain in the frequency band to down-convert the input RF signal.

Finally,the frequency response of the filter is tested in four paths,as shown in Fig.7.The input signal power is about -34 dBm,and LO is set at 2 GHz.In the figure,the light blue line,yellow line,pink line,and blue line represent the frequency response curves of the 200 MHz,400 MHz,800 MHz,and 1.5 GHz filter

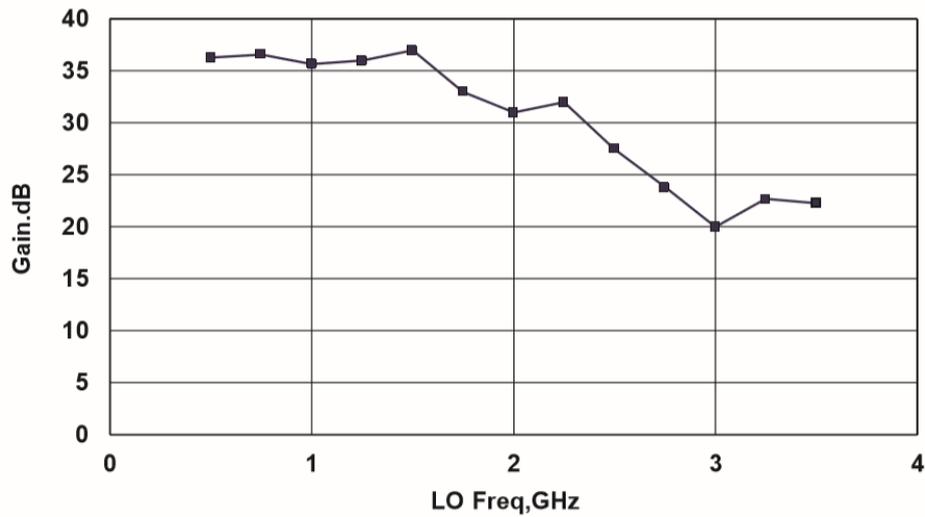

Fig. 6: The variation of the mixing path gain under the different LO settings.

paths,respectively.A total of 7 frequency points in the 200 MHz and 400 MHz filter paths are measured,and 10 frequeny points are measured in the 800 MHz and 1500 MHz filter paths.

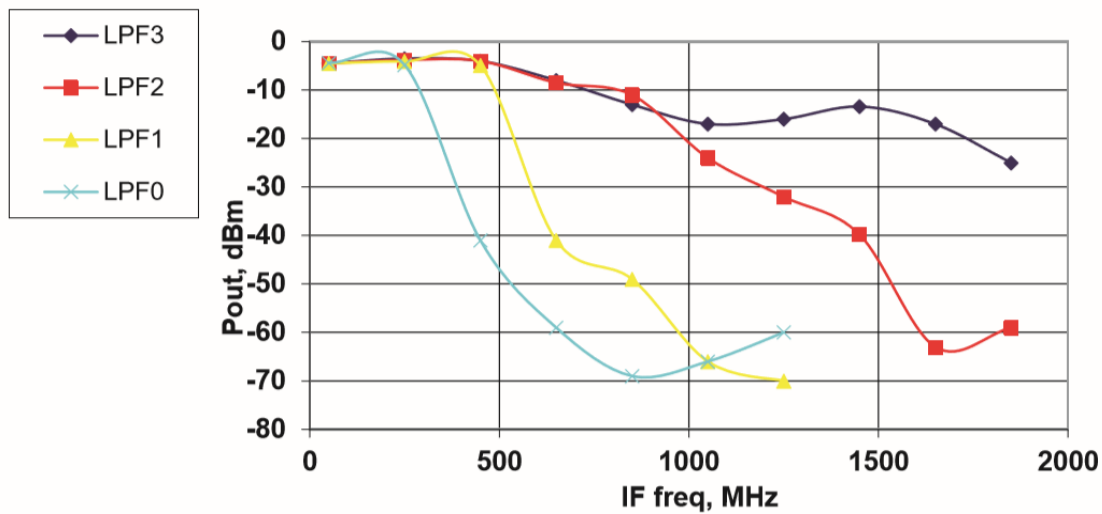

Fig. 7: Frequency response of four paths.

## 3  WIDEBAND HIGH PRECISION ANALOG DIGITAL CONVERTER

The Analog Digital Converter acquisition section is composed of ADC acquisition hardware and ADC ac-quisition control logic.The ADC acquisition hardware refers to the ADC acquisition circuit that can realize 6 Gsps sampling rate,and the ADC acquisition logic indicates the FPGA logic that controls the acquisition part. The AD acquisition part consists of two ADC12D1600 chips with a sampling rate of 3 GHz. Two ADs are sampled in parallel by the synchronization mechanism, to realize the overall sampling rate of 6 GHz. The AD chip uses the ADC12D1600. The overall AD parallel sampling structure block diagram is as follows.For each ADC, the clock signal is 1.5GHz, and the 1a and 1b of the chip ADC1 collect the data on the rising and falling edges of the 1.5GHz sampling clock respectively to achieve the 3GHz sampling. For 1a and 1b in an ADC chip, the sampling clock is equivalent to a 0-degree offset clock and a 180-degree offset clock of a 1.5 GHz clock,respectively.To achieve a total sample rate of 6 GHz, the two ADCs are synchronized and the clocks of the two ADC inputs are phase shifted by 90 degrees. Let the input sampling clock of ADC2 be a 90-degree phase-shifted clock of 1.5 GHz. In this case, the sampling clock of 2a is a 90-degree offset clock of 1.5 GHz, and 2b is a 270-degree phase-shifted clock of 1.5 GHz.The sampling rate of 6 GHz is achieved.

### 3.1 Synchronization mechanism

This ADC12D1600 chip has two functions,namely,automatic synchronization and DCLK reset,so that users can synchronize multiple ADCs in the system.The automatic synchronization function specifies that one chip is the primary ADC,and the other is the slave ADC.For applications with multiple primary and slave ADCs,automatic synchronization is used to synchronize the slave ADCs from their respective primary ADCs,while the primary ADCs are synchronized by the reset of DCLK.Automatic synchronization is a new feature that continuously synchronizes the output of multiple ADC12D1600s in the system.It synchro-nizes DCLK of ADC and data outputs with a primary ADC.Its advantage is that no special synchronization pulse is needed.All synchronous timing errors can be recovered in the next DCLK cycle.The master-slave ADCs are connected in the form of binary trees,so that the synchronization errors can be flushed out of the system quickly.Automatic synchronization synchronizes the DCLK clock of each ADC to load the in-put data into the FPGA .The RCOUT interface of the slave ADC is connected through the control register configuration ADCs to set the master ADC and the slave ADC.DCLK is used to synchronize multiple pri-mary ADCs.However, our project only needs two ADCs,and one master ADC is included.Therefore,DCLK reset function is not needed. Timestamp is a feature of the ADC12D1600 chip,and it allows the user to input a synchronization signal to the Time Stamp.The effective bit number of the 12 bit ADC is generally 9 bits,while timestamp occupies the lowest bit D0 of the whole 12 bits.The function will not affect the perfor-mance of the ADC. An identical timestamp signal is sent to multiple ADCs.Since each ADC can receive the input signal,the data stream of the ADC output can be aligned according to the edge of timestamp.Based on this feature,the delay of the output data stream between the two ADCs can be corrected.In other words,the delay between the timestamp signals output by the two ADCs represents the delay between the output data signals.According to the information,the data received by the two ADCs can be aligned. Aperture delay helps to eliminate the mismatches between ADCs resulted from different signal paths or clock offset.The size of the aperture delay can be configured through the ADC configuration registers.It can be modified in real time based on the debug situation to find a final balance value.

### 3.2 Clocking device

Two sampling clocks with a phase offset of 90 degrees are provided by a dedicated clock board which also outputs timestamp signals to the system for synchronization.The LMK04808 EVALUATION board is selected as the clock board,as shown in Fig.8.

### 3.3 Hardware testing of the wideband high performance ADC

In order to test the hardware performance of the FAST3212 ADC,24 frequency signals are sent to the ADC separately at the power of 0 dBm.The SNR at each frequency point is calculated,as shown in Fig.10.At 10 MHz point,the SNR is about 84 dB.As the frequency increases gradually,the SNR of the system decreases.When the signal frequency is 3.2 GHz,the SNR decreases to 70 dB.The ADC sampling principle determines whether the strength of the input signal is fixed and discrete.The SINC function is actually ex-pressed as roll off.Meanwhile,the coupler is used for the sampling circuit board of the ADC chip,which acts as a low-pass filter.Superposition will cause the attenuation of signal strength,leading to the loss of SNR in calculation.In fact,the signal strength will be adjusted for compensation.

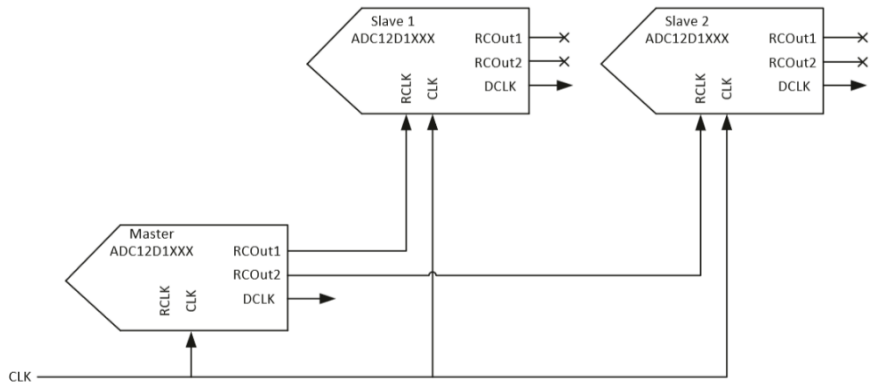

Fig. 8: AutoSync Example.

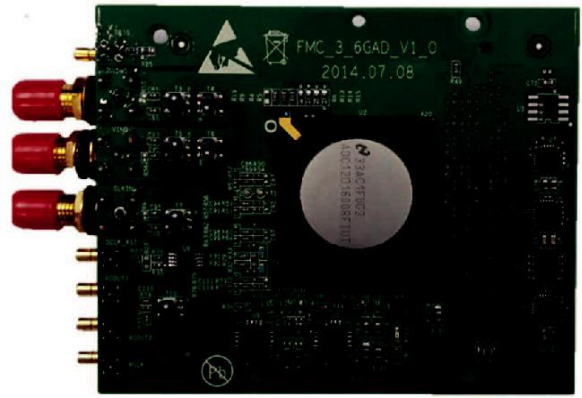

Fig. 9: ADC12D1600 12 bit ADC.

A total of 14 input signals of different powers are selected at the signal frequency of 156.25 MHz,and the SNR corresponding to each power point is calculated.When the power of the signal is -30 dBm,the SNR is 58 dB.However,with the increase of the signal power,the SNR of the corresponding signal increases linearly until 85 dB,and this power is the maximum value of ADC(1.2 VPP or 5.5 dBm).When the power of the input signal exceeds the maximum value of ADC,the ADC is overloaded and the noise component increases significantly.Consequently,the SNR gradually decreases.

## 4 FAST DIGITAL BACK-END

### 4.1 Hardware architecture of FDB

The original intention of the FDB system development is to implement a large point FFT using a cascade algorithm.The digital back-end adopts the overall architecture of 6U standard size FPGA motherboard with FMC VITA 57.1 standard AD sub-board and high-speed protocol tailboard.The computing board is equipped with a virtex-5 core(xc5vlx50t-1ff665) as the master chip. Two virte

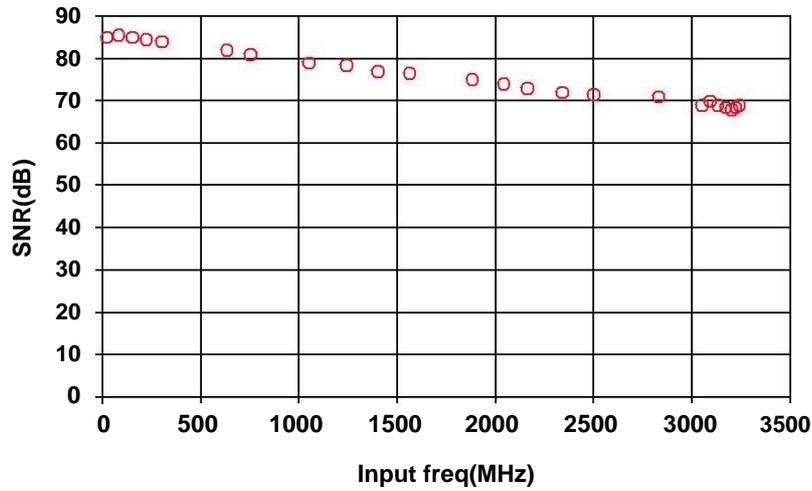

Fig. 10: SNR at 0dBm.

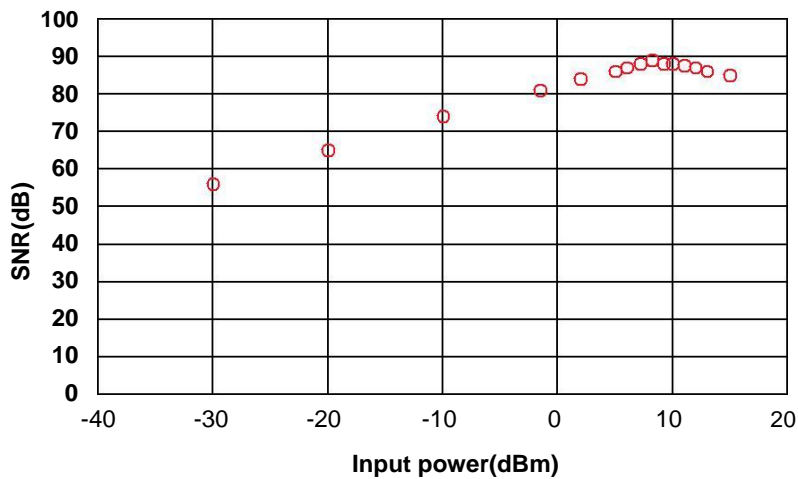

Fig. 11: SNR at 156.25MHz.

chips(xc6vlx240t-2ff1759) are used as its computing chipXia et al. (2016).The N-point FFT is divided into two layers.The first layer has 65536 points,and the second layer has 128 points.There are up to 8 million channels in the two computing chips by the cascade algorithms.The sophisticated spectral structure has good resolution. External storage includes 2GB DDR3 and 9 M QDR.At least 12 gigabit Ethernet ports are supported on the transport tail board.

   Another SNAP2 high performance hardware platform is shown in Fig.13.A Kintex ultra-scale series high performance FPGA is equipped as the main processor,including 5520 DSP multiplier and 2160 36Kb BRAM.A piece of Zynq7000 SOC co-processor includes dual cortex-a9 ARM and Artix7 FPGA logic re-sources.Two pieces of high precision FAST3212 ADC can be inserted to realize 2-way acquisition.External storage includes 1GB DDR3,144Mbit QDR.Four QSFP interfaces at the front end are used for high-speed data transmission,The parameters are downloaded and controlled by two Ethernet

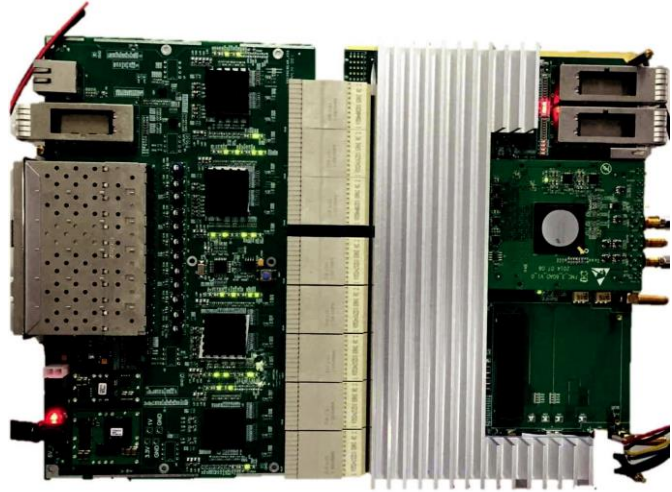

Fig. 12: FDB.

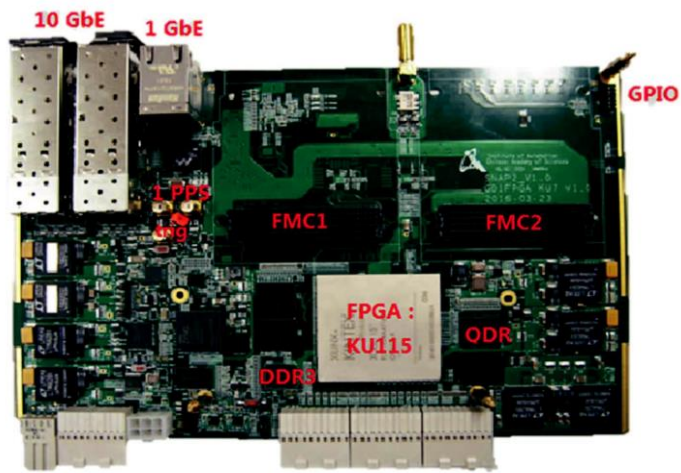

Fig. 13: SNAP2.

Table 1: The details of different working modes in CRANE system.

| No | Mode | Sample rate | Bandwidth | Output bits | File format |
|---|---|---|---|---|---|
| 1 | Narrow spectral line | 6GS/s | 31.25 MHz | 32 float | SDFITS |
| 2 | Incoherent pulsar | 4GS/s | 2000 MHz | 8 bits | PSRFITS |
| 3 | Baseband | 4GS/s | 250 MHz | 8 bits | Raw data |

interfaces,and up to seven QSFP interfaces can be extended at the back end through the tail plate for data transmission.

The observation modes of CRANE include broad spectral line mode,narrow spectral line mode,incoherent pulsar mode and baseband data storage mode.The detailed information of different work-ing modes is listed in Table 1.

The narrow spectral line mode is selected 31.25 MHz,configurable mixer,output channels are 512 k,the channel bandwidth is approximately 0.0596 kHz,full stokes polarizations.The processing flow chart is shown in Fig.14.The working mode of ADC can be configured through ACC MON module,and the number of ADC is taken to examine whether the data is correct to realize the monitoring function of ADC module.After that, the data is stored through FIFO,packaged and sent to the High Performance Compute node.The actual data transmission rate of the network is 1 GB/s.

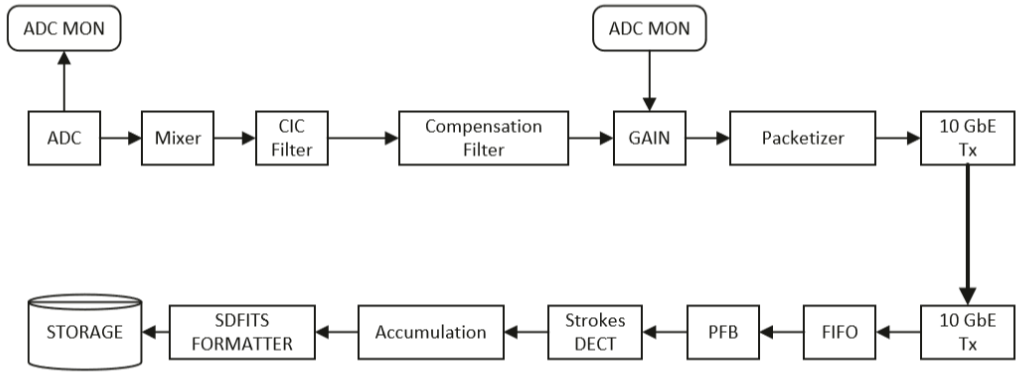

Fig. 14: Processing flow chart of narrow bandwidth mode

In the incoherent pulsar mode,the number of the output channels can be selected as 2K,4K,8K,16K,32k and 64k.The dump rate is about 16 microseconds and 50 microseconds at 2K channel and 64K chan-nel,respectively. The processing flow chart of incoherent pulsar mode is shown in Fig.15.

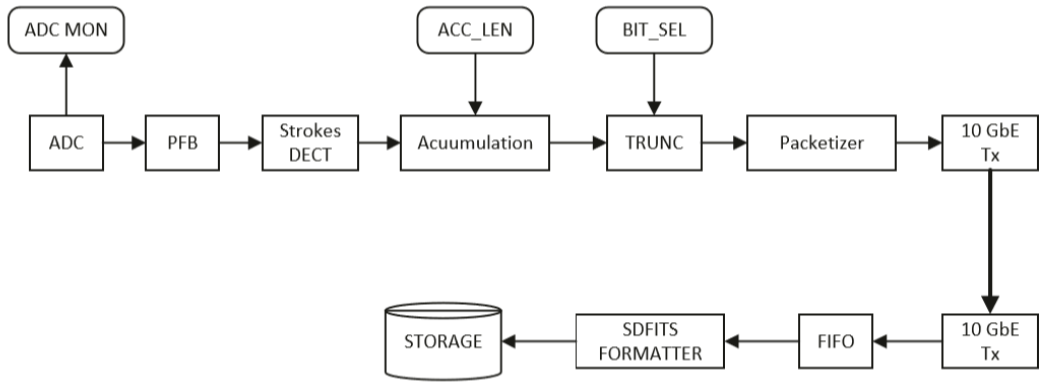

Fig. 15: Processing flow chart of incoherent pulsar mode.eps

The baseband mode is selected 250 MHz,configurable mixer.Through the Decimation Filter,the bandwidth output needed for observation can be chosen.The processing flow chart is shown in Fig.16.

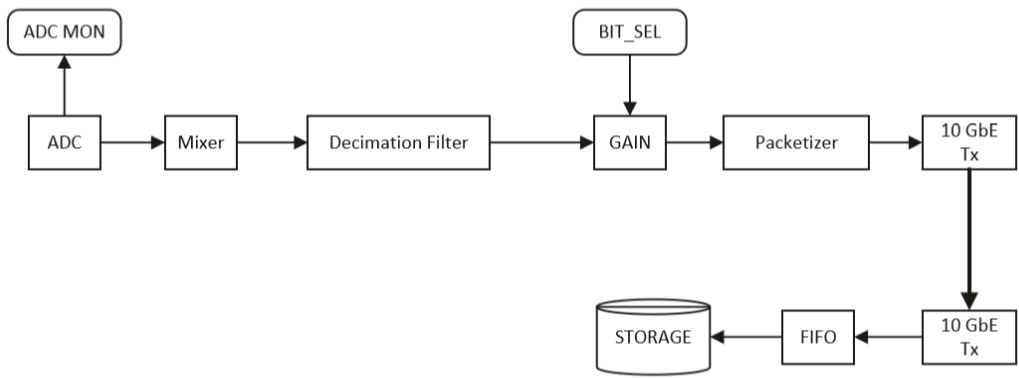

Fig. 16: Baseband mode

Table 1: The details of different working modes in CRANE system.

| No | Mode | Sample rate | Bandwidth | Output bits | File format |
|---|---|---|---|---|---|
| 1 | Narrow spectral line | 6GS/s | 31.25 MHz | 32 float | SDFITS |
| 2 | Incoherent pulsar | 4GS/s | 2000 MHz | 8 bits | PSRFITS |
| 3 | Baseband | 4GS/s | 250 MHz | 8 bits | Raw data |

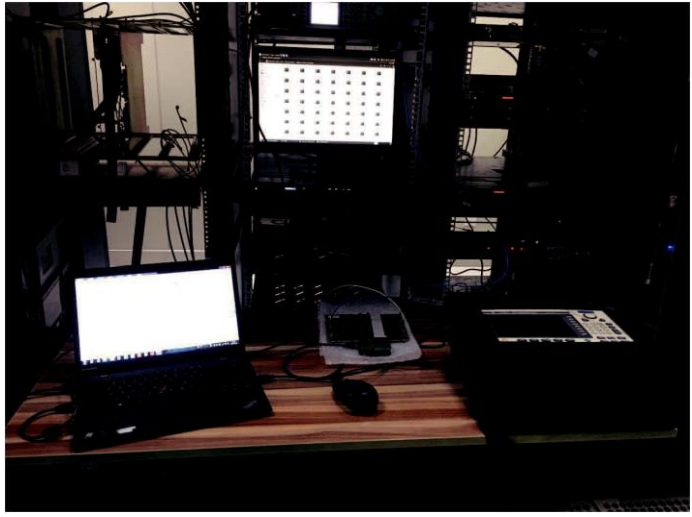

Fig. 17: The photo of CRANE working in one observation at FAST site

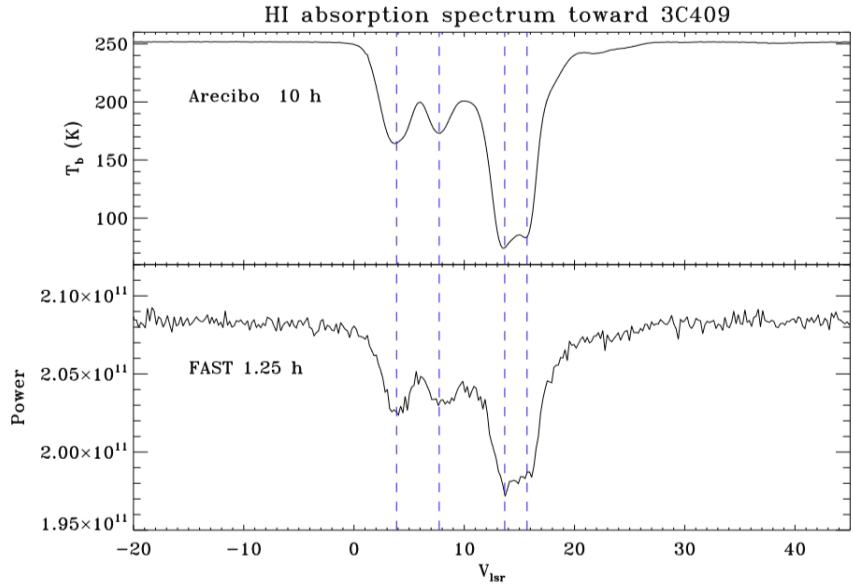

Fig. 18: Absorption spectrum toward 3C409. Results from Arecibo and FAST are shown in upper and lower panel, respectively.

## 5 OBSERVATION

### 5.1 HI absorption line

When we using CRANE for observation the working photo is shown in Fig.17.FAST ultra-wideband receiver detected the neutral hydrogen absorption line of the 3C409 in drifting scan mode on August 14,2017.Pointing observations are carried out toward the quasar 3C409, the flux density of which is 13.6

Jy at 1.4 GHz. ON and 4 OFF positions with the offset of 4.5 arcmin are observed. As shown in Fig.18, the structure of the derived absorption spectrum is consistent with that from Arecibo.

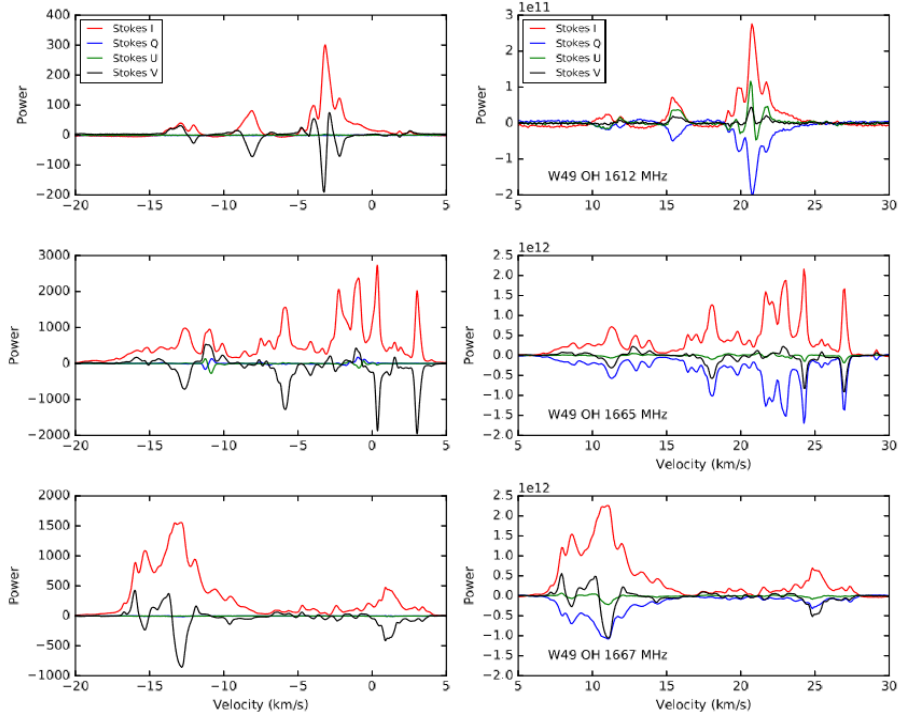

Fig.19: The comparison of the Arecibo calibrated polarization spectra and the FAST uncalibrated polariza tion spectra of W49 OH masers.The three figures on the left are the results of the Arecibo telescope,and the three figures on the right are the results of the FAST telescope.Since the calibration can't be performed at that time,the unit is the intensity of the CRANE output,by comparing the figures on both sides,we can find that the contours of the two are consistent,which is enough to illustrate the reliability of the spectral lines.

## 5.2 Polarization

Fig.19 describes the comparisons of the Arecibo calibrated polarization spectra and the FAST uncalibrated polarization spectra of W49 OH masers at 1612 (top panels), 1665 (middel panels), and 1667 (bottom panels) MHz. The FAST circular polarization (Stokes V) at OH 1667 MHz is consistent with the result of Arecibo.The FAST circular polarization at OH 1665 MHz is weaker than that of Arecibo.The FAST circular polarization at OH 1612 MHz has an opposite sign to that of Arecibo. The Stokes Q of FAST is much larger than the calibrated Stokes Q of Arecibo, which is a natural outcome for the linear polarization feed of FAST ultrawide-band receiver.

## 6 CONCLUSIONS

From the beginning of 2017 to the middle of 2018,CRANE was the only spectral line observation digital backend in FAST site.After completing all spectral line commissioning and observation tasks,the results were consistent with the expectations.Benefited from the high sensitivity of the telescope,the CRANE sys-tem needs less observation time to achieve the same spectral line intensity compared with the Arecibo.Based on the spectral line observations of CRANE,multiple conference reports,articles(to be determined) and project results were produced.The technology adopted in the CRANE system will also be utilized in the projects requested by High Energy Institute.In addition,the CRANE system can continuously improve the observation performance of FRB and SETI,and observe FRB and SETI as early as possible.Meanwhile,we will cooperate with the electromagnetic working group to enhance the performance of achieving RFI miti-gation technology,and use the cascade FFT algorithm to obtain more channels for adaptive filters.